\documentclass[amsmath,amssymb,pre,preprint]{revtex4-1}
\usepackage{graphicx}
\usepackage{dcolumn}
\usepackage{bm}
\usepackage{mathtools}
\usepackage{graphicx}
\usepackage{pictexwd}
\usepackage{dcolumn}
\usepackage{amsmath}
\usepackage[usenames,dvipsnames]{xcolor}
\newcommand{\synopsis}[1]{}
\newcommand{\noopsort}[1]{}

\newcommand{\oldversion}[1]{}

\begin{document}


\title{Debye-H\"uckel Theory of Weakly Curved Macroions: Implementing Ion Specificity through a Composite Coulomb-Yukawa Interaction Potential}

\author{Bjorn K.~Berntson}
\affiliation{Department of Mathematics, North Dakota
  State University, Fargo ND  58108, USA}

\author{Rachel Downing}
\affiliation{Department of Physics, North Dakota
  State University, Fargo ND 58108, USA}
 
\author{Guilherme Volpe Bossa}
\affiliation{Department of Physics, North Dakota
  State University, Fargo ND 58108, USA}

\author{Sylvio May} \email{sylvio.may@ndsu.edu}
\affiliation{Department of Physics, North Dakota
  State University, Fargo ND  58108, USA}

\date{\today}

\begin{abstract}
The free energy of a weakly curved, isolated macroion embedded in a symmetric 1:1 electrolyte solution is calculated on the basis of linear Debye-H\"uckel theory, thereby accounting for non-electrostatic Yukawa pair interactions between the mobile ions and of the mobile ions with the macroion surface, present in addition to the electrostatic Coulomb potential. The Yukawa interactions between anion-anion, cation-cation, and anion-cation pairs are independent from each other and serve as a model for solvent-mediated ion-specific effects. We derive expressions for the free energy of a planar surface, the spontaneous curvature, the bending stiffness, and the Gaussian modulus. It is shown that a perturbation expansion, valid if the Yukawa interactions make a small contribution to the overall free energy, yields simple analytic results that exhibit good agreement with the general free energy over the range of experimentally relevant interaction parameters.
\end{abstract}


\maketitle

\section{Introduction}
Debye-H\"uckel theory accounts for the influence of Coulomb interactions between the ions of an electrolyte in the dilute limit. Its foundation is the description of the ionic atmosphere using the linearized Poisson-Boltzmann equation. When applied to individual ions of a bare and uniform electrolyte, the theory is able to rationalize measured ion activity coefficients at very low electrolyte concentrations \cite{debye23,wright07}. Another line of application is the description of the electric double layer (EDL) near weakly charged macroions such as proteins \cite{gilson88}, lipid membranes \cite{pincus90}, microemulsions \cite{levine72}, and polyelectrolytes \cite{micka96}. It also has been used extensively to describe interactions between these macroions \cite{schiessel98,kunze00,cherstvy07}.  

The Debye-H\"uckel model relies on significant approximations such as the neglect of ion shape, polarizability, hydration, and spatial variations of the dielectric constant. The model nevertheless has significant appeal because it is simple, linear, serves as the (often analytically accessible) dilute limit of classical Poisson-Boltzmann theory, and can be used to develop extensions \cite{ben-yaakov09,bohinc2017}. One of these extensions is to complement the Coulomb interaction by an additional non-electrostatic pair potential to describe soft, solvent-mediated hydration interactions \cite{stafiej98,burak_2000,burak01,ruckenstein_2002,paillusson10}. These additional interactions are a means to incorporate ion specificity \cite{manciu03,kunz10,ben11b} into the modeling of the EDL \cite{brown15}. The Yukawa potential has received special attention \cite{bohinc11,bohinc12,buyukdagli11,buyukdag12,zhao12,bandopadhyay13}, despite the fact that molecular-level simulations suggest solvent-mediated ion-ion potentials exhibit an osciallatory component \cite{kalcher09,kalcher10}. Recent work has presented the systematic incorporation of independent Yukawa-like anion-anion, cation-cation, and anion-cation interactions in addition to the Coulomb potential \cite{caetano16,caetano17} and compared the predictions of mean-field theory with Monte Carlo simulations. Analytic solutions of the full nonlinear mean-field model are not available, not even for a single planar surface. However it is feasible (and, in fact, one of the goals of the present work) to derive analytic expressions for the free energy in the Debye-H\"uckel regime.

When charged surfaces in an electrolyte are curved, the EDL undergoes a spatial reorganization. The corresponding free energy change can be expressed in terms of a set of curvature elastic constants that have been calculated previously within the framework of Debye-H\"uckel \cite{winterhalter88} and nonlinear Poisson-Boltzmann \cite{lekkerkerker89a,lekkerkerker90,winterhalter92} theory. However, predictions of the curvature elastic properties in the presence of composite Coulomb-Yukawa pair interactions have not been investigated so far. We therefore include the analysis of weakly curved surfaces into this work.

We calculate the free energy of a weakly curved macroion embedded in a symmetric 1:1 electrolyte in the limit of linearized Debye-H\"uckel electrostatics, where ion-ion and ion-surface interactions derive from composite Coulomb-Yukawa pair potentials. While the Coulomb potential describes the electrostatic properties of the EDL, the Yukawa component serves as a convenient model for ion specificity. More specifically, two anions separated by a distance $r$ interact through the electrostatic potential $l_B/r$, where $l_B$ is the Bjerrum length and where here and in the following, all interaction potentials and energies are expressed in units of the thermal energy $k_BT$ (Boltzmann's constant $k_B$ times the absolute temperature $T$). The corresponding expressions for two cations and an anion-cation pair are $l_B/r$ and $-l_B/r$, respectively. Note that $l_B=0.7 \: \mbox{nm}$ in an aqueous solution at room temperature. In addition to that, ions also interact through Yukawa potentials: $\bar{a} e^{-\kappa (r-\bar{a})}/r$ for two anions, $\bar{b} e^{-\kappa (r-|\bar{b}|)}/r$ for an anion-cation pair, and $\bar{c} e^{-\kappa (r-\bar{c})}/r$ for two cations. Here $1/\kappa$ is a characteristic decay length that is set by the structure of the solvent, and the constants $\bar{a}$, $\bar{b}$, $\bar{c}$ determine the Yukawa interaction strengths. We have defined these constants in analogy to the Bjerrum length. That is, the Yukawa interaction between two anions is equal to the thermal energy unit if their mutual distance is $r=\bar{a}$, and similarly for two cations ($r=\bar{c}$), and anion-cation pairs ($r=|\bar{b}|$). Note that in the latter case we use the absolute value $|\bar{b}|$ because $\bar{b}$ may adopt negative values, whereas we demand $\bar{a}$ and $\bar{c}$ to be non-negative due to symmetry. It will be convenient to re-express the interaction strengths as $a=\bar{a} e^{\kappa \bar{a}}$, $b=\bar{b} e^{\kappa |\bar{b}|}$, and $c=\bar{c} e^{\kappa \bar{c}}$, so that the three Yukawa potentials read $a e^{-\kappa r}/r$, $b e^{-\kappa r}/r$, and $c e^{-\kappa r}/r$. Our model also includes solvent-induced ion-surface interactions; they emerge naturally as boundary conditions of the differential equations that describe our composite Coulomb-Yukawa interactions.

\section{Classical Debye-H\"uckel Theory} \label{lo45}
For an electrolyte of uniform dielectric constant that contains monovalent salt ions of bulk concentration $n_0$, electrostatic interactions can be described by a dimensionless potential $\Psi_e=\Psi_e({\bf r})$ that satisfies the Poisson equation $\nabla^2 \Psi_e=4 \pi l_B (n_a-n_c)$. Here, ${\bf r}$ denotes a position within the electrolyte, $n_a=n_a({\bf r})$  the local anion concentration, and $n_c=n_c({\bf r})$ the local cation concentration. Note that $\Psi_e=e \Phi/k_BT$ is related to the electrostatic potential $\Phi=\Phi({\bf r})$, where $e$ denotes the elementary charge. According to the classical Poisson-Boltzmann model, anions and cations are Boltzmann-distributed, $n_a=n_0 e^{\Psi_e}$ and $n_c=n_0 e^{-\Psi_e}$. This leads to the classical Poisson-Boltzmann equation, $l_D^2 \nabla^2 \Psi_e=\sinh \Psi_e$ or, in the linearized Debye-H\"uckel regime, $l_D^2 \nabla^2 \Psi_e=\Psi_e$, where $l_D=(8 \pi l_B n_0)^{-1/2}$ denotes the Debye screening length. A macroion with fixed (but not necessarily uniform) surface charge density $\sigma_e$ is associated with the boundary condition $(\partial \Psi_e/\partial n)_s=-4 \pi l_B \sigma_e/e$, where $(\partial/\partial n)_s$ denotes the derivative in the normal direction of the macroion surface, pointing into the electrolyte. The index ``s'' indicates that the derivative is taken at the macroion surface. If the macroion is isolated, the potential and its gradient must vanish far away from the macroion. These two boundary conditions fully define the potential $\Psi_e({\bf r})$ for any macroion geometry. The surface potential can be used to compute the free energy of the EDL that forms in the vicinity of the macroion. On the level of linear Debye-H\"uckel theory, the free energy is $F=(1/2) \int do \Psi_e \sigma_e/e$, where the integration runs over the entire macroion surface. 

Weakly curved macroions have local radii of curvature much larger than the Debye screening length $l_D$. In this case, we can Helfrich-expand \cite{helfrich73} the free energy per unit area $A$
\begin{equation} \label{jp69}
\frac{F}{A}=\frac{F_0}{A}+\frac{k}{2} (c_1+c_2)^2-k c_0 (c_1+c_2)+\bar{k} c_1 c_2,
\end{equation}
where $c_1$ and $c_2$ denote the two principal curvatures, $F_0$ the free energy of a planar surface, $k$ the bending stiffness, $\bar{k}$ the Gaussian modulus, and $c_0$ the spontaneous curvature. Note that stability of a surface that is allowed to curve requires $2 k>-\bar{k}>0$. The classical Debye-H\"uckel model yields \cite{winterhalter88} $F_0/A=2 \pi l_B l_D (\sigma_e/e)^2$, $k c_0=\pi l_B l_D^2 (\sigma_e/e)^2$, $k=(3/2) \pi l_B l_D^3 (\sigma_e/e)^2$, and $\bar{k}=-(2/3) k$. These results, which also appear as the small $\sigma_e$-limit of the predictions for the non-linear Poisson-Boltzmann theory \cite{lekkerkerker89a}, account only for Coulomb pair-interactions between all involved charge carriers (mobile ions and charges on the macroion surface). No non-electrostatic interactions, such as excluded volume effects or hydration forces among the mobile ions and between the mobile ions and the macroion surface, are accounted for.

In the following we generalize the results of the classical Debye-H\"uckel model to the presence of a composite Coulomb-Yukawa pair-potential.

\section{Ion-Specific Debye-H\"uckel Theory}
As outlined in the Introduction, we assume that solvent-mediated hydration interactions can be described in terms of the ion-specific pair-potentials $a e^{-\kappa r}/r$ for two anions, $b e^{-\kappa r}/r$ for an anion-cation pair, and $c e^{-\kappa r}/r$ for two cations. Similarly to the Coulomb interaction that can be expressed in terms of an electrostatic potential $\Psi_e({\bf r})$ which fulfills Poisson's equation, the hydration interactions give rise to two potentials $\Psi_a({\bf r})$ and $\Psi_c({\bf r})$ which fulfill the Helmholtz equations
\begin{equation}\label{nn2}
\begin{pmatrix}
(\nabla^2-\kappa^2) \: \Psi_a({\bf r})\\(\nabla^2-\kappa^2) \: \Psi_c({\bf r})
\end{pmatrix}=-4 \pi \: \mathcal{A}\begin{pmatrix}
n_a({\bf r})-n_0  \\ n_c({\bf r})-n_0 \end{pmatrix}
\end{equation}
with complex wavenumber and a source term. Note that $\Psi_a({\bf r})$ and $\Psi_c({\bf r})$ are defined relative to the bulk, where $n_a=n_c=n_0$. Hence, in the bulk $\Psi_a=\Psi_c=0$. The matrix
\begin{equation}\label{jo48}
\mathcal{A}=\begin{pmatrix}
a & b  \\ b & c 
\end{pmatrix}
\end{equation}
describes the interaction strengths. The origin of Eqs.~\ref{nn2} and \ref{jo48} is discussed in Appendix I and in Caetano {\em et al} \cite{caetano16}. We note that two hydration potentials are needed in the most general case where the determinant of $\mathcal{A}$ does not vanish. As introduced above, the parameters $a=\bar{a} e^{\kappa \bar{a}}$, $b=\bar{b} e^{\kappa |\bar{b}|}$, and $c=\bar{c} e^{\kappa \bar{c}}$, describe the strengths of the Yukawa pair potentials, $a$ for an anion-anion pair, $b$ for an anion-cation pair, and $c$ for a cation-cation pair. Symmetry demands $a \ge 0$ and $c \ge 0$, whereas $b$ may adopt positive or negative values. Recall that the anion-anion Yukawa interaction is equal to the thermal energy unit for $r=\bar{a}$, and analogously for anion-cation pairs ($r=|\bar{b}|$) and for cation-cation pairs ($r=\bar{c}$).

Minimization of an appropriate mean-field free energy (see Appendix I for details) that accounts for the composite Coulomb-Yukawa pair potential in addition to ideal mixing contributions of the ions yields the Boltzmann distributions \cite{caetano16}
\begin{equation}\label{ik32}
n_a=n_0 \: e^{\Psi_e-\Psi_a}, \hspace{2cm} n_c=n_0 \: e^{-\Psi_e-\Psi_c}.
\end{equation}
Inserting these into the Poisson and Helmholtz equations leads to a set of three non-linear differential equations for the three potentials
\begin{eqnarray}\label{er43}
\nabla^2 \Psi_e&=&4 \pi l_B n_0 \left(e^{\Psi_e-\Psi_a}-e^{-\Psi_e-\Psi_c} \right),\nonumber\\
\begin{pmatrix}
(\nabla^2-\kappa^2) \: \Psi_a({\bf r})\\(\nabla^2-\kappa^2) \: \Psi_c({\bf r})
\end{pmatrix}&=&4 \pi n_0 \: \mathcal{A} \begin{pmatrix}
 1-e^{\Psi_e-\Psi_a} \\ 1-e^{-\Psi_e-\Psi_c} \end{pmatrix}.
\end{eqnarray}
These equations generalize the classical Poisson-Boltzmann model to the additional presence of Yukawa interactions. In the absence of these (for $\Psi_a=\Psi_c=0$), Eqs.~\ref{er43} recover the classical Poisson-Boltzmann equation $l_D^2 \nabla^2 \Psi_e=\sinh \Psi_e$ with $l_D=(8 \pi l_B n_0)^{-1/2}$.

In the following, we focus exclusively on the Debye-H\"uckel limit, which corresponds to the linearization of Eqs.~\ref{er43}, valid if all three potentials are sufficiently small,
\begin{eqnarray}\label{bm90}
\nabla^2 \Psi_e&=&\frac{1}{l_D^2} \Psi_e+\frac{1}{2 l_D^2} \left(\Psi_c-\Psi_a\right),\nonumber\\
\nabla^2 \Psi_a-\kappa^2 \Psi_a&=&\frac{1}{l_a^2} \left(-\Psi_e+\Psi_a\right)+\frac{1}{l_b^2} \left(\Psi_e+\Psi_c\right),\\
\nabla^2 \Psi_c-\kappa^2 \Psi_c&=&\frac{1}{l_b^2} \left(-\Psi_e+\Psi_a\right)+\frac{1}{l_c^2} \left(\Psi_e+\Psi_c\right),\nonumber
\end{eqnarray}
where we have defined $l_a$, $l_b$ and $l_c$ through
\begin{equation} \label{ki51}
  \frac{1}{l_a^2}=4 \pi a n_0, \hspace{0.3cm}
  \frac{1}{l_b^2}=4 \pi b n_0, \hspace{0.3cm}
  \frac{1}{l_c^2}=4 \pi c n_0.
\end{equation}
We observe this system of differential equations is invariant under switching the identity of anions and cations (which includes charge inversion): $\Psi_a \leftrightarrow  \Psi_c$, $\Psi_e \rightarrow -\Psi_e$, $l_a \leftrightarrow l_c $. Eqs.~\ref{bm90} can be cast into the more compact form,
\begin{equation} \label{fr52}
l_D^2 \nabla^2 {\bf \Psi}={\cal B} {\bf \Psi},
\end{equation}
expressed in terms of the column vector ${\bf \Psi}=\left(\Psi_e,\Psi_a,\Psi_c\right)$ and the matrix
\begin{equation}\label{zr39}
{\cal B}=l_D^2
\begin{pmatrix}
  \frac{1}{l_D^2} & -\frac{1}{2 l_D^2} & \frac{1}{2 l_D^2} \\
-\frac{1}{l_a^2}+\frac{1}{l_b^2} & \frac{1}{l_a^2}+\kappa^2 & \frac{1}{l_b^2} \\
-\frac{1}{l_b^2}+\frac{1}{l_c^2} & \frac{1}{l_b^2} & \frac{1}{l_c^2}+\kappa^2
\end{pmatrix}.
\end{equation}
We assume the macroion carries a fixed surface charge density $\sigma_e$. In addition, we also allow for solvent-mediated interactions of the mobile ions with the macroion surface, expressed by the two parameters $\sigma_a$ and $\sigma_c$ that we cast into the column vector $\boldsymbol{\sigma}=\left(\sigma_e/e,\sigma_a,\sigma_c\right)$. Similarly to $\sigma_e/e$ being the surface density of the sources for the Coulomb interaction, $\sigma_a$ and $\sigma_c$ characterize the surface density of the sources for the ion-surface Yukawa interactions: $\sigma_a$ for the anions and $\sigma_c$ for the cations. 
(At this point we regard $\sigma_e/e$, $\sigma_a$, and $\sigma_c$ as a set of fixed thermodynamic variables that reflect electrode properties and that our curvature-expanded free energy depends on. Of course, we are free to---and below will -- introduce couplings between $\sigma_e/e$, $\sigma_a$, and $\sigma_c$).
Note that $\sigma_a$ and $\sigma_c$ can adopt positive or negative values. If $a$, $b$, $c$, $\sigma_e$ and $\sigma_c$ are all positive, the macroion surface repels all mobile ions. If $a$, $b$, $c$, $-\sigma_e$ and $-\sigma_c$ are all positive, the macroion surface attracts all mobile ions. The choice $a=c=-b$ and $\sigma_e=\sigma_c$ leaves the macroion surface inert. In the general case, the boundary condition for solving Eq.~\ref{fr52} can be written as
\begin{equation} \label{fr57}
l_D {\left(\frac{\partial {\bf \Psi}}{\partial n}\right)}_s=-{\cal M} \boldsymbol{\sigma},
\end{equation}
where $(\partial/\partial n)_s$ denotes the derivative in the normal direction of the macroion surface, pointing into the electrolyte. Also, in Eq.~\ref{fr57} we have defined the matrix
\begin{equation}\label{fy48}
\mathcal{M}=4 \pi l_D \begin{pmatrix}
l_B & 0 & 0  \\ 0 & a & b \\ 0 & b  & c
\end{pmatrix}.
\end{equation}
For an isolated macroion we demand that all three potentials, $\Psi_e$, $\Psi_a$, $\Psi_c$, and their gradients vanish far away from the macroion.

We model solvent-mediated interactions on the basis of Yukawa potentials. It is reasonable to assume solvent is present only outside the macroion but not inside. This case corresponds to the interaction strength of the Yukawa potential being zero inside the macroion. The boundary condition in Eq.~\ref{fr57} therefore only contains contributions from the fields outside the macroion. If an aqueous solvent (or a solvent of different type) was present inside the macroion, the fields $\Psi_a$ and $\Psi_c$ (more specifically, their derivatives at the macroion surface taken into the normal direction pointing inside the macroion) would contribute to the boundary condition. We do not consider this case in the present work.

\section{Free Energy Calculation for Weakly Curved Macroion}
The free energy of an isolated macroion corresponding to the ion-specific Debye-H\"uckel model can be calculated (see Appendix I) according to 
\begin{equation}\label{hj42}
F=\frac{1}{2} \int do \left[\frac{\sigma_e}{e} \Psi_e+\sigma_a \Psi_a+\sigma_c \Psi_c\right]=\frac{1}{2} \int do  \boldsymbol{\Psi} \cdot \boldsymbol{\sigma}.
\end{equation}
If $\boldsymbol{\sigma}$ is fixed at the macroion surface, then what we need in order to execute the calculation of $F$ is the dependence of $\boldsymbol{\Psi}$ on $\boldsymbol{\sigma}$. Our goal is to compute that dependence and, from that, an explicit expression for the free energy of a single, isolated, weakly curved macroion. The term ``weakly curved'' refers to radii of curvature that are much larger than any of the characteristic lengths $l_D$, $l_a$, $l_b$, and $l_c$ (we take $|l_b|$ if $b<0$). In this case we can, again, Helfrich-expand the free energy per unit area $A$, as specified in Eq.~\ref{jp69}. This reduces our goal to the calculation of the free energy for a planar surface $F_0$, the spontaneous curvature $c_0$, the bending stiffness $k$, and the Gaussian modulus $\bar{k}$. To this end, we re-express Eq.~\ref{fr52} for cylindrical ($n=1$), and spherical ($n=2$) symmetry,
\begin{equation}\label{zd57}
  l_D^2 \left[\frac{d^2 \boldsymbol{\Psi}}{d r^2}+\frac{n}{r} \frac{d \boldsymbol{\Psi}}{d r}\right]={\cal B} \boldsymbol{\Psi},
\end{equation}
where $r$ is the corresponding radial coordinate of a cylindrical or spherical coordinate system. We introduce a new dimensionless distance $x$ (with $x \ge 0$) via $r=1/c+x l_D$, where $c-c_1=c_2=0$ for cylindrical and $c=c_1=c_2$ for spherical geometry. Note that $x$ measures the scaled distance from the weakly curved macroion surface to a position within the EDL. For our potentials we write up to second order in curvature $\boldsymbol{\Psi}(x)=\boldsymbol{\Psi}_0(x)+c l_D \boldsymbol{\Psi}_1(x)+c^2 l_D^2 \boldsymbol{\Psi}_2(x)$. Expanding Eq.~\ref{zd57} up to second order in $c$ yields three linear equations for the three curvature-components $\boldsymbol{\Psi}_0(x)$, $\boldsymbol{\Psi}_1(x)$, and $\boldsymbol{\Psi}_2(x)$, 
\begin{eqnarray}\label{zs21}
  \boldsymbol{\Psi}_0''&=&{\cal B} \boldsymbol{\Psi}_0,\nonumber\\
  \boldsymbol{\Psi}_1''+n \boldsymbol{\Psi}_0'&=&{\cal B} \boldsymbol{\Psi}_1,\\
  \boldsymbol{\Psi}_2''+n \boldsymbol{\Psi}_1'-n x \boldsymbol{\Psi}_0'&=&{\cal B} \boldsymbol{\Psi}_2.\nonumber 
\end{eqnarray}
We can carry out a first integration subject to the boundary condition that all potentials, $\boldsymbol{\Psi}_0(x)$, $\boldsymbol{\Psi}_1(x)$, $\boldsymbol{\Psi}_2(x)$, (and their derivatives) vanish in the limit $x \rightarrow \infty$,
\begin{eqnarray}\label{aw44}
  \boldsymbol{\Psi}_0'&=&-{\cal B}^{1/2} \boldsymbol{\Psi}_0,\nonumber\\
  \boldsymbol{\Psi}_1'&=&-{\cal B}^{1/2} \boldsymbol{\Psi}_1-\frac{n}{2}\boldsymbol{\Psi}_0,\\
  \boldsymbol{\Psi}_2'&=&-{\cal B}^{1/2} \boldsymbol{\Psi}_2-\frac{n}{2}\boldsymbol{\Psi}_1+\frac{n}{2} x \boldsymbol{\Psi}_0+\frac{n}{4} \left(1-\frac{n}{2}\right) {\cal B}^{-1/2} \boldsymbol{\Psi}_0.\nonumber 
\end{eqnarray}
Note that ${\cal B}^{1/2}$ is defined such that ${\cal B}^{1/2} {\cal B}^{1/2}={\cal B}$, and ${\cal B}^{-1}$ denotes the inverse of ${\cal B}$ such that  ${\cal B}^{-1} {\cal B}$ yields the identity matrix. The boundary condition in Eq.~\ref{fr57} imposes fixed surface densities for $\sigma_e/e$, $\sigma_a$, and $\sigma_c$, independent of curvature. This implies $\boldsymbol{\Psi}_0'(x=0)=-{\cal M} \boldsymbol{\sigma}$ and $\boldsymbol{\Psi}_1'(x=0)=\boldsymbol{\Psi}_2'(x=0)=\boldsymbol{0}$, with the column vector $\boldsymbol{0}=\left(0,0,0\right)$. Using these boundary conditions and applying Eqs.~\ref{aw44} to the macroion surface, $x=0$, gives rise to a linear system of equations for the curvature components of the surface potential. Solving this linear system provides us with the explicit expressions 
\begin{eqnarray}\label{ix19}
  \boldsymbol{\Psi}_0(0)&=&{\cal B}^{-1/2} {\cal M} \boldsymbol{\sigma},\nonumber\\
  \boldsymbol{\Psi}_1(0)&=&-\frac{n}{2} {\cal B}^{-1} {\cal M} \boldsymbol{\sigma},\\
\boldsymbol{\Psi}_2(0)&=&\frac{n}{4} \left(\frac{n}{2}+1\right) {\cal B}^{-3/2} {\cal M} \boldsymbol{\sigma},\nonumber  
\end{eqnarray}
for how the surface potential $\boldsymbol{\Psi}(0)=\boldsymbol{\Psi}_0(0)+c l_D \boldsymbol{\Psi}_1(0)+c^2 l_D^2 \boldsymbol{\Psi}_2(0)$ depends on the surface densities $\boldsymbol{\sigma}$. If we insert $\boldsymbol{\Psi}(0)$ into Eq.~\ref{hj42}, both for cylindrical ($n=1$) and for spherical ($n=2$) curvature, and compare with the corresponding expressions, $F/A=F_0/A+k c^2/2-k c_0 c$ for cylindrical symmetry ($n=1$) and $F/A=F_0/A+(2 k+\bar{k}) c^2-2 k c_0 c$ for spherical symmetry ($n=2$), we find
\begin{eqnarray} \label{lp08}
\frac{F_0}{A}&=&\frac12 \boldsymbol{\sigma}^T {\cal B}^{-1/2} {\cal M} \boldsymbol{\sigma}, \nonumber  \\
k c_0&=&\frac{l_D}{4}\boldsymbol{\sigma}^T {\cal B}^{-1} {\cal M} \boldsymbol{\sigma}, \\
k&=&\frac{3}{8} l_D^2 \boldsymbol{\sigma}^T {\cal B}^{-3/2} {\cal M} \boldsymbol{\sigma}, \nonumber  \\
\bar{k}&=&-\frac{2}{3} k. \nonumber  
\end{eqnarray}
where $\boldsymbol{\sigma}^T$ is the transpose of $\boldsymbol{\sigma}$. Eq.~\ref{lp08} is the principal result of the present work. As expected on the level of Debye-H\"uckel theory, the expressions in Eq.~\ref{lp08} are quadratic forms of the surface densities $\sigma_e/e$, $\sigma_a$, and $\sigma_c$. These quadratic forms represent general results of a weakly curved macroion (with fixed $\sigma_e/e$, $\sigma_a$, and $\sigma_c$) in the presence of a composite Coulomb-Yukawa pair interaction. Recall the matrix ${\cal M}$ is specified in Eq.~\ref{fy48}, and the matrix ${\cal B}$ in Eq.~\ref{zr39}. Regarding the latter, recall the definitions $l_a$, $l_b$, and $l_c$ in Eq.~\ref{ki51}.  To obtain explicit expressions for $F_0$, $c_0$, $k$, and $\bar{k}$ in terms of the interaction parameters $a$, $b$, $c$, $\kappa$, and the salt concentration $n_0$, we need to find ${\cal B}^{-1/2}$, ${\cal B}^{-1}$, and ${\cal B}^{-3/2}$. This can easily be accomplished numerically for any given set of system parameters.
 
\section{Discussion}
Bazant {\em et al} \cite{bazant11} have recently suggested a phenomenological approach to account for short-range correlations among ions, leading to a term $\sim \nabla^4 \Psi_e$ contained in a generalized nonlinear Poisson-Boltzmann equation. Using theories of binary fluid mixtures, a similar fourth-order Poisson-Boltzmann equation was derived by Blossey {\em et al} \cite{blossey17}. Our present approach, which requires us to introduce the two additional fields $\Psi_a$ and $\Psi_c$ in order to account for independent Yukawa anion-anion, anion-cation, and cation-cation interactions, leads to a {\em sixth-order} differential equation for the electrostatic potential $\Psi_e$. On the Debye-H\"uckel level that equation is a linear one. Specifically, from Eq.~\ref{bm90} we find
\begin{equation} \label{lo96}
  \nabla^6 \Psi_e-C_4 \nabla^4 \Psi_e+C_2 \nabla^2 \Psi_e=C_0 \Psi_e
\end{equation}
with the coefficients
\begin{eqnarray} \label{ft78}
  C_0&=&\frac{\kappa^2}{2 l_D^2} \left[2 \kappa^2+\frac{1}{l_a^2}+\frac{2}{l_b^2}+\frac{1}{l_c^2}\right],\nonumber\\
  C_2&=&\kappa^4-\frac{1}{l_b^4}+\kappa^2 \left(\frac{1}{l_a^2}+\frac{2}{l_b^2}+\frac{1}{l_c^2}\right)\\
  &+&\frac{1}{2} \left(\frac{2}{l_b^2 l_D^2}+\frac{1}{l_c^2 l_D^2}+\frac{1}{l_a^2 l_D^2}+\frac{1}{l_a^2 l_c^2} \right),\nonumber \\
  C_4&=&2 \kappa^2+\frac{1}{l_a^2}+\frac{1}{l_c^2}+\frac{1}{l_D^2}.\nonumber 
\end{eqnarray} 
Combinations of exponential solutions with three characteristic lengths will emerge from Eq.~\ref{lo96}; they depend on $\kappa$, $l_D$, $l_a$, $l_b$, and $l_c$.

An analytic calculation of ${\cal B}^{-1/2}$, ${\cal B}^{-1}$, and ${\cal B}^{-3/2}$ yields cumbersome expressions. However, a few specific cases lead to simple results and thus to meaningful explicit expressions for $F_0$, $c_0$, $k$, and $\bar{k}$. We discuss those in the following.

\subsection{Symmetric Yukawa Interactions}
The first specific case is $a=b=c$, where all ions, irrespective of being anions or cations, interact with each other through the same Yukawa potential. Eqs.~\ref{lp08} then give rise to
\begin{eqnarray} \label{hg64}
  \frac{F_0}{A}&=&2 \pi l_B l_D {\left(\frac{\sigma_e}{e}\right)}^2+2 \pi a
  \frac{(\sigma_a+\sigma_c)^2}{\sqrt{\kappa^2+2/l_a^2}},\nonumber\\
k c_0&=&\pi l_B l_D^2 {\left(\frac{\sigma_e}{e}\right)}^2+\pi a \frac{(\sigma_a+\sigma_c)^2}{\kappa^2+2/l_a^2},\\
k&=&\frac{3}{2} \pi l_B l_D^3 {\left(\frac{\sigma_e}{e}\right)}^2+\frac{3}{2} \pi a \frac{(\sigma_a+\sigma_c)^2}{(\kappa^2+2/l_a^2)^{3/2}},\nonumber
\end{eqnarray} 
and $\bar{k}/k=-2/3$, as before. Clearly, the curvature-dependent free energy decomposes into additive Coulomb and Yukawa contributions. The two contributions act independently, without any coupling. The first contribution to $F_0/A$, $k c_0$, $k$ in Eq.~\ref{hg64} is identical to the result of the classical Debye-H\"uckel model as stated in Sec.~\ref{lo45}. The second contribution reflects the presence of particles that are uniformly distributed on a surface with area density $\sigma_a+\sigma_c$ and exhibit mutual Yukawa interactions $a e^{-\kappa_{\mathrm{eff}} r}/r$. Here, $\kappa_{\mathrm{eff}}=\sqrt{\kappa^2+2/l_a^2}$ is an effective inverse screening length that differs from $\kappa$ because of the interaction of the salt ions (which are present with a combined bulk concentration of $2 n_0$) with the surface. For example, the Yukawa contribution to the free energy (per unit area) of a planar surface amounts to 
\begin{equation} \label{pw91}
\frac{F_0}{A}=2 \pi a (\sigma_a+\sigma_c)^2 \int \limits_0^\infty dr r \frac{e^{-\kappa_{\mathrm{eff}} r}}{r}=2 \pi a \frac{(\sigma_a+\sigma_c)^2}{\sqrt{\kappa^2+2/l_a^2}},  
\end{equation}
which recovers the Yukawa contribution in the first line of Eq.~\ref{hg64}. The Yukawa contributions to $k c_0$ and $k$ in Eq.~\ref{hg64} follow from a similar calculation.
For our discussion below we also note that for sufficiently small $a=b=c \rightarrow \delta a$, Eqs.~\ref{hg64} read
\begin{eqnarray} \label{hh66}
\frac{F_0}{A}&=&2 \pi l_B l_D {\left(\frac{\sigma_e}{e}\right)}^2+\frac{2 \pi}{\kappa} (\sigma_a+\sigma_c)^2 \delta a,\nonumber\\
k c_0&=&\pi l_B l_D^2 {\left(\frac{\sigma_e}{e}\right)}^2+\frac{\pi}{\kappa^2} (\sigma_a+\sigma_c)^2 \delta a,\\
k&=&\frac{3}{2} \pi l_B l_D^3 {\left(\frac{\sigma_e}{e}\right)}^2+\frac{3}{2} \: \frac{\pi}{\kappa^3} (\sigma_a+\sigma_c)^2 \delta a.\nonumber
\end{eqnarray} 
Here, the Yukawa contribution acts as a small perturbation for the result from the classical Debye-H\"uckel model.

\subsection{Perturbation approach}
The second specific case starts from the classical Debye-H\"uckel model and introduces the parameters $a$, $b$, $c$ as first-order perturbations. In this case, we can express Eqs.~\ref{lp08} as the sum of a pure electrostatic contribution plus a perturbation due to non-vanishing (but small) parameters $a \rightarrow \delta a$, $b \rightarrow \delta b$, and $c \rightarrow \delta c$, 
\begin{eqnarray} \label{lp98}
  \frac{F_0}{A}&=&2 \pi l_B l_D {\left(\frac{\sigma_e}{e}\right)}^2+
  \frac{1}{2} \boldsymbol{\sigma}^T \delta[{\cal B}^{-1/2} {\cal M}] \boldsymbol{\sigma},\nonumber\\
k c_0&=&\pi l_B l_D^2 {\left(\frac{\sigma_e}{e}\right)}^2+
  \frac{1}{4} \boldsymbol{\sigma}^T \delta[{\cal B}^{-1} {\cal M}] \boldsymbol{\sigma},\\
k&=&\frac{3}{2} \pi l_B l_D^3 {\left(\frac{\sigma_e}{e}\right)}^2+
  \frac{3}{8} \boldsymbol{\sigma}^T \delta[{\cal B}^{-3/2} {\cal M}] \boldsymbol{\sigma}.\nonumber
\end{eqnarray}
The perturbation contributions amount to (see Appendix II for details) 
\begin{eqnarray} \label{gt54}
  \delta[{\cal B}^{-1/2} {\cal M}]&=&\frac{2 \pi}{\kappa} \left(\begin{array}{ccc}
    \frac{\delta a+\delta c-2 \delta b}{g_1}  & \frac{\delta a-\delta b}{g_2} & \frac{\delta b-\delta c}{g_2}  \\
  \frac{\delta a-\delta b}{g_2}  & 2 \delta a & 2 \delta b \\  \frac{\delta b-\delta c}{g_2} & 2 \delta b & 2 \delta c \\ \end{array}\right),
\nonumber\\
\delta[{\cal B}^{-1} {\cal M}]&=&\frac{2 \pi}{\kappa^2} \left(\begin{array}{ccc}
\frac{\delta a+\delta c-2 \delta b}{g_3}  & \frac{\delta a-\delta b}{g_4} & \frac{\delta b-\delta c}{g_4}  \\
  \frac{\delta a-\delta b}{g_4}  & 2 \delta a & 2 \delta b \\  \frac{\delta b-\delta c}{g_4} & 2 \delta b & 2 \delta c \\ \end{array}\right),
\\
\delta[{\cal B}^{-3/2} {\cal M}]&=&\frac{2 \pi}{\kappa^3} \left(\begin{array}{ccc}
\frac{\delta a+\delta c-2 \delta b}{g_5} &  
\frac{\delta a-\delta b}{g_6} & \frac{\delta b-\delta c}{g_6}  \\
\frac{\delta a-\delta b}{g_6} & 2 \delta a & 2 \delta b \\  
\frac{\delta b-\delta c}{g_6} & 2 \delta b & 2 \delta c \\ \end{array}\right),\nonumber
\end{eqnarray} 
where we define $g_1=4 (1+\tilde{\kappa})^2/(2+\tilde{\kappa})$, $g_2=1+\tilde{\kappa}$, $g_3=2$, $g_4=1$, $g_5=[4 (1+\tilde{\kappa})^2]/\{2+\tilde{\kappa} [4+3 \tilde{\kappa} (2+\tilde{\kappa})]\}$, $g_6=(1+\tilde{\kappa})/[1+\tilde{\kappa}+\tilde{\kappa}^2]$, and $\tilde{\kappa}=\kappa l_D$. As expected, for $\delta a=\delta b=\delta c$ the expressions in Eqs.~\ref{lp98} and \ref{gt54} become identical to those in Eq.~\ref{hh66}. In the general case of asymmetric Yukawa interactions ($\delta a+\delta c \neq 2 \delta b$) electrostatic and Yukawa interactions are coupled. For example, the specific case $\sigma_a=\sigma_c=0$ implies that 
\begin{eqnarray} \label{cq95}
  \frac{F_0}{A}=2 \pi \left[l_B l_D+\frac{\delta a+\delta c-2 \delta b}{2 \kappa g_1} \right] \: {\left(\frac{\sigma_e}{e}\right)}^2,\nonumber\\
k c_0=\pi \left[l_B l_D^2+\frac{\delta a+\delta c-2 \delta b}{2 \kappa^2 g_3} \right] \: {\left(\frac{\sigma_e}{e}\right)}^2,\\
k=\frac{3}{2} \pi \left[l_B l_D^3+\frac{\delta a+\delta c-2 \delta b}{2 \kappa^3 g_5} \right] \: {\left(\frac{\sigma_e}{e}\right)}^2\nonumber
\end{eqnarray}
all grow (for $\delta a+\delta c > 2 \delta b$) or decrease (for $\delta a+\delta c < 2 \delta b$), when the Yukawa interactions are switched on.

The perturbation contribution for the free energy of a planar macroion surface in Eq.~\ref{cq95}, 
\begin{equation}
\frac{\delta F_0}{A}=\frac{2+\tilde{\kappa}}{4 (1+\tilde{\kappa})^2} \: \frac{\pi}{\kappa} \: {\left(\frac{\sigma_e}{e}\right)}^2 (\delta a+\delta c-2 \delta b),
\end{equation}
assumes $\sigma_a=\sigma_c=0$. In the following, we analyze the general case where $\sigma_e$, $\sigma_a$, $\sigma_c$ may all be non-vanishing. In principle, $\sigma_e$, $\sigma_a$, and $\sigma_c$ are independent parameters that reflect electrode properties. A convenient way to discuss the behavior of $F_0/A$, $k c_0$, and $k$ for general choices of $\sigma_e$, $\sigma_a$, and $\sigma_c$ is to couple the solvent-induced ion-surface interactions $\sigma_a=\chi_a \sigma_e/e$ and $\sigma_c=\chi_c \sigma_e/e$ to the electrostatic surface charge density $\sigma_e$, where $\chi_a$ and $\chi_c$ are two dimensionless coupling parameters. That is, instead of using $\sigma_e$, $\sigma_a$, $\sigma_c$ we use the set $\sigma_e$, $\chi_a$, $\chi_c$ as independent variables. We point out that $\chi_a$ and $\chi_c$ are auxiliary quantities that merely facilitate the systematic discussion (in the remainder of this subsection) of Eqs.~\ref{lp98} and \ref{gt54}. Of course, for any specific choice of $\sigma_e$, $\chi_a$, and $\chi_c$, the actual thermodynamic variables $\sigma_e$, $\sigma_a$, and $\sigma_c$ follow immediately.

\begin{figure}[ht!]
\begin{center}
  \includegraphics[width=8.0cm]{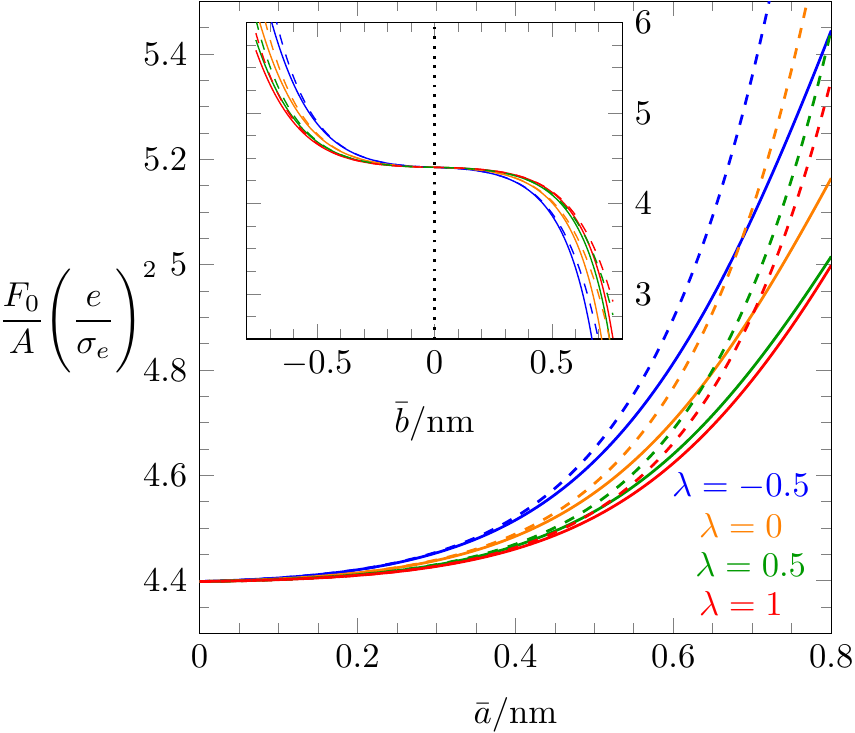}
\end{center}
\caption{\label{fig1} 
Scaled free energy, $F_0/A \times (e/\sigma_e)^2$ (in units of $k_BT \mbox{nm}^2$), of a planar surface as function of $\bar{a}$ for $l_B=0.7 \: \mbox{nm}$, $l_D=1 \: \mbox{nm}$, $1/\kappa=0.2 \: \mbox{nm}$, $\bar{b}=\bar{c}=0$. Solid lines refer to the full result in Eq.~\ref{lp08}, broken lines to the perturbation result in Eq.~\ref{lp98}. Curves of different color correspond to different couplings, $\chi_a=\lambda \chi_a^{\mathrm{opt}}$ and $\chi_c=\lambda \chi_c^{\mathrm{opt}}$, with $\chi_c^{\mathrm{opt}}=-\chi_a^{\mathrm{opt}}=1/(2 g_2)=0.0833$, and $\lambda=-0.5$ (blue), $\lambda=0$ (orange, vanishing coupling), $\lambda=0.5$ (green), $\lambda=1$ (red, optimal coupling). The inset shows $F_0/A \times (e/\sigma_e)^2$ as function of $\bar{b}$ for $\bar{a}=\bar{c}=0$, with otherwise the same parameters and color code as in the main diagram.
}
\end{figure}

The two coupling parameters, $\chi_a$ and $\chi_c$, can be optimized by requiring $\partial F_0/\partial \chi_a=0$ and $\partial F_0/\partial \chi_c=0$. This gives rise to $\chi_a=\chi_a^{\mathrm{opt}}$ and $\chi_c=\chi_c^{\mathrm{opt}}$ with $\chi_c^{\mathrm{opt}}=-\chi_a^{\mathrm{opt}}=1/(2 g_2)=1/[2 (1+\tilde{\kappa})]$. Note that $\chi_a^{\mathrm{\mathrm{opt}}}<0$ and $\chi_c^{\mathrm{opt}}>0$. Hence, at optimal coupling and for both $a>b$ and $c>b$, when the surface becomes positively charged ($\sigma_e>0$), with anions accumulating and cations depleting from the surface, the anions experience an additional non-electrostatic attraction to the surface (because of $\chi_a^{\mathrm{opt}}<0$), and the cations experience an additional non-electrostatic repulsion from the surface (because of $\chi_c^{\mathrm{opt}}>0$). Upon inserting $\chi_a=\chi_a^{\mathrm{opt}}$ and $\chi_c=\chi_c^{\mathrm{opt}}$, we obtain for the perturbation contribution of the free energy
\begin{equation}
\frac{\delta F_0}{A}=\frac{\tilde{\kappa}}{4 (1+\tilde{\kappa})^2} \: \frac{\pi}{\kappa} \: {\left(\frac{\sigma_e}{e}\right)}^2 (\delta a+\delta c-2 \delta b).
\end{equation}
We point out that using the optimal coupling parameters $\chi_a^{\mathrm{opt}}$ and $\chi_c^{\mathrm{opt}}$ in the free energy corresponds to fixing the surface potentials $\Psi_a(x=0)=\Psi_c(x=0)=0$ when changing $\sigma_e$. The ratio between the free energy perturbations for vanishing coupling and optimal coupling is 
\begin{equation} \label{lr54}
\frac{\delta F_0(\chi_a=0,\chi_c=0)}{\delta F_0(\chi_a=\chi_a^{\mathrm{opt}},\chi_c=\chi_c^{\mathrm{opt}})}=\frac{2+\tilde{\kappa}}{\tilde{\kappa}}>1.
\end{equation}
To illustrate this result we show in the main diagram of Fig.~\ref{fig1} the scaled free energy $F_0/A \times (e/\sigma_e)^2$ according to Eq.~\ref{lp08} (solid lines, the full result) and Eq.~\ref{lp98} (broken lines, the perturbation result), calculated for $l_B=0.7 \: \mbox{nm}$, $l_D=1 \: \mbox{nm}$, $1/\kappa=0.2 \: \mbox{nm}$, $\bar{b}=\bar{c}=0$, and plotted as function of $\bar{a}$. 
Note that plotting the scaled free energy as function of $\bar{a}$ (instead of $a=\bar{a} e^{\kappa \bar{a}}$) is meaningful because for an anion-anion distance $r=\bar{a}$ the hydration interaction for that ion pair amounts to the thermal energy $k_BT$. Curves of different color in Fig.~\ref{fig1} correspond to different couplings $\chi_a=\lambda \chi_a^{\mathrm{opt}}$ and $\chi_c=\lambda \chi_c^{\mathrm{opt}}$ with $\chi_c^{\mathrm{opt}}=-\chi_a^{\mathrm{opt}}=1/(2 g_2)$ and $\lambda=-0.5$ (blue), $\lambda=0$ (orange, vanishing coupling), $\lambda=0.5$ (green), and $\lambda=1$ (red, optimal coupling). As predicted by Eq.~\ref{lr54}, the change of $F_0$ becomes minimal for optimal coupling but does not change its sign. The same reasoning is also true when $\bar{b}$ or $\bar{c}$ are changed instead of $\bar{a}$. This is illustrated in the inset of Fig.~\ref{fig1}, which shows $F_0/A \times (e/\sigma_e)^2$ as function of $\bar{b}$ with $\bar{a}=\bar{c}=0$, for otherwise the same parameters and color code as in the main diagram. 

\begin{figure}[ht!]
\begin{center}
\includegraphics[width=8.0cm]{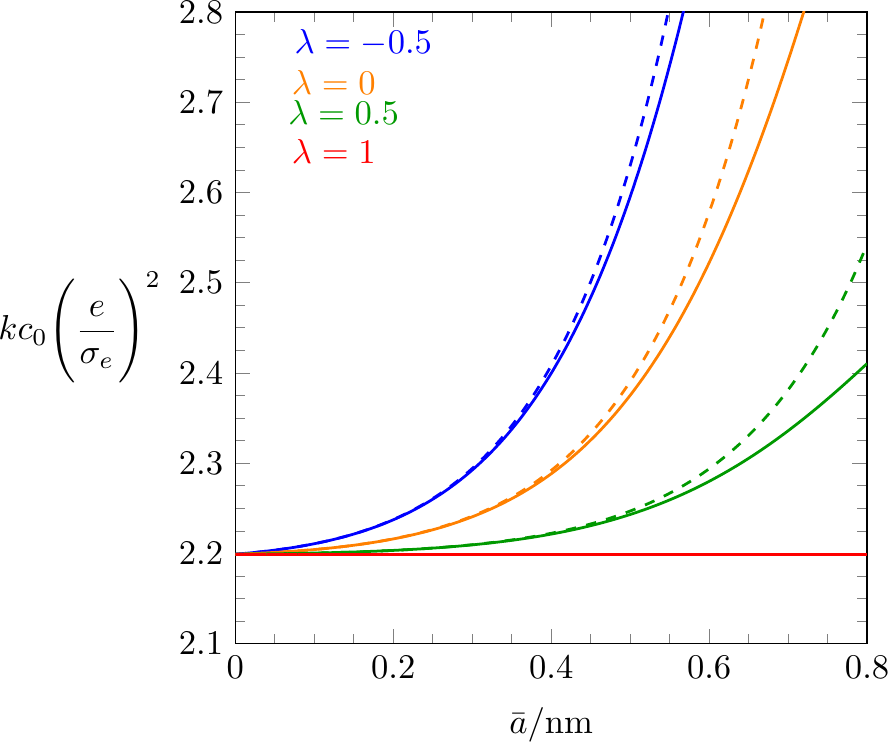}
\end{center}
\caption{\label{fig2} 
Scaled bending stiffness times spontaneous curvature, $k c_0 \times (e/\sigma_e)^2$ (in units of $k_BT \mbox{nm}^3$), of a planar surface as function of $\bar{a}$ for $l_B=0.7 \: \mbox{nm}$, $l_D=1 \: \mbox{nm}$, $1/\kappa=0.2 \: \mbox{nm}$, $\bar{b}=\bar{c}=0$. Solid lines refer to the full result in Eq.~\ref{lp08}, broken lines to the perturbation result in Eq.~\ref{lp98}. Curves of different color correspond to different couplings, $\chi_a=\lambda \chi_a^{\mathrm{\mathrm{opt}}}$ and $\chi_c=\lambda \chi_c^{\mathrm{\mathrm{opt}}}$, with $\chi_c^{\mathrm{opt}}=-\chi_a^{\mathrm{opt}}=1/(2 g_4)=1/2$, and $\lambda=-0.5$ (blue), $\lambda=0$ (orange, vanishing coupling), $\lambda=0.5$ (green), $\lambda=1$ (red, optimal coupling). 
}
\end{figure}

A similar calculation can be carried out for the perturbation contribution to the term $k c_0$ in Eqs.~\ref{lp98} and \ref{gt54}. We again define the two coupling parameters $\chi_a$ and $\chi_c$ through $\sigma_a=\chi_a \sigma_e/e$ and $\sigma_c=\chi_c \sigma_e/e$. At optimal coupling ($\chi_a=\chi_a^{\mathrm{opt}}$ and $\chi_c=\chi_c^{\mathrm{opt}}$) these two parameters fulfill the relations $\partial (k c_0)/\partial \chi_a=0$ and $\partial (k c_0)/\partial \chi_c=0$, implying $\chi_c^{\mathrm{opt}}=-\chi_a^{\mathrm{opt}}=1/(2 g_4)=1/2$. At optimal coupling we find a vanishing spontaneous curvature contribution, $\delta (k c_0)=0$. Hence, any nonvanishing coupling between $\sigma_e/e$, $\sigma_a$, and $\sigma_c$ can reduce the magnitude of the spontaneous curvature perturbation but not change its sign. We illustrate this in Fig.~\ref{fig2}, which shows $k c_0 \times (e/\sigma_e)^2$ according to Eq.~\ref{lp08} (solid lines, the full result) and Eq.~\ref{lp98} (broken lines, the perturbation result) for the same parameters as in Fig.~\ref{fig1}.
As predicted by the perturbation result, at optimal coupling (the red curve in Fig.~\ref{fig2}) there is no change in spontaneous curvature when $\bar{a}$ is switched on. 

\begin{figure}[ht!]
\begin{center}
  \includegraphics[width=8.0cm]{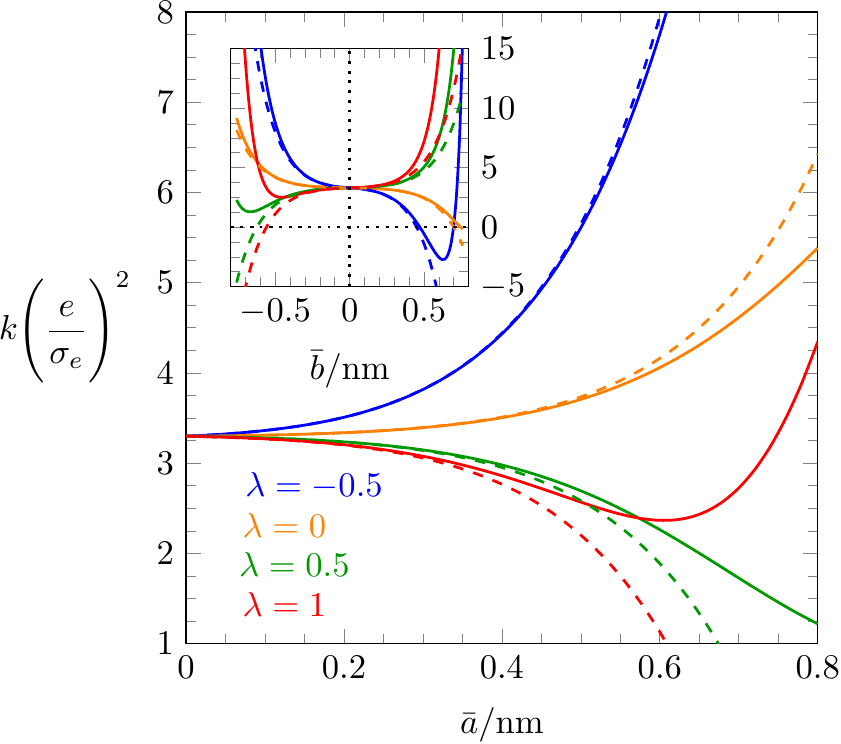}
\end{center}
\caption{\label{fig3} 
Scaled bending stiffness, $k \times (e/\sigma_e)^2$ (in units of $k_BT \mbox{nm}^4$), of a planar surface as function of $\bar{a}$ for $l_B=0.7 \: \mbox{nm}$, $l_D=1 \: \mbox{nm}$, $1/\kappa=0.2 \: \mbox{nm}$, $\bar{b}=\bar{c}=0$. Solid lines refer to the full result in Eq.~\ref{lp08}, broken lines to the perturbation result in Eq.~\ref{lp98}. Curves of different color correspond to different couplings, $\chi_a=\lambda \chi_a^{\mathrm{opt}}$ and $\chi_c^{\mathrm{opt}}=-\chi_a^{\mathrm{opt}}=1/(2 g_6)=2.58$, and $\lambda=-0.5$ (blue), $\lambda=0$ (orange, vanishing coupling), $\lambda=0.5$ (green), $\lambda=1$ (red, optimal coupling). The inset shows $k \times (e/\sigma_e)^2$ as function of $\bar{b}$ for $\bar{a}=\bar{c}=0$, with otherwise the same parameters and color code as in the main diagram.
}
\end{figure}

Finally, for the perturbation contribution to the bending stiffness in Eqs.~\ref{lp98} and \ref{gt54} we again introduce $\chi_a$ and $\chi_c$ as before and determine the optimal coupling $\chi_a=\chi_a^{\mathrm{opt}}$ and $\chi_c=\chi_c^{\mathrm{opt}}$ from $\partial k/\partial \chi_a=0$ and $\partial k/\partial \chi_c=0$. This yields $\chi_c^{\mathrm{opt}}=-\chi_a^{\mathrm{opt}}=1/(2 g_6)=(1+\tilde{\kappa}+\tilde{\kappa}^2)/[2 (1+\tilde{\kappa})]$ and using that, 
\begin{equation}
\frac{\delta k(\chi_a=0,\chi_c=0)}{\delta k(\chi_a=\chi_a^{\mathrm{opt}},\chi_c=\chi_c^{\mathrm{opt}})}=-\frac{2}{\tilde{\kappa}^3}-\frac{6}{\tilde{\kappa}}+\frac{9}{1+2 \tilde{\kappa}},
\end{equation}
which is negative for all $\tilde{\kappa}>0$. Hence, upon changing the coupling
parameters from zero to $\chi_a^{\mathrm{opt}}$ and $\chi_c^{\mathrm{opt}}$, the sign of the perturbation contribution $\delta k$ must change. This is illustrated in Fig.~\ref{fig3}, which shows $k \times (e/\sigma_e)^2$ according to Eq.~\ref{lp08} (solid lines, the full result) and Eq.~\ref{lp98} (broken lines, the perturbation result) for the same parameters as in Figs.~\ref{fig1} and \ref{fig2}. 
Clearly, the sign of the change in bending stiffness as function of $\bar{a}$ switches when the coupling parameters $\chi_a$ and $\chi_c$ are changed from zero (the orange curves in Figs.~\ref{fig3}) to their optimal values $\chi_a^{\mathrm{opt}}$ and $\chi_c^{\mathrm{opt}}$ (the red curves in Figs.~\ref{fig3}). We also observe this when changing $\bar{b}$ instead of $\bar{a}$; see the inset of Fig.~\ref{fig3}. Moreover, the inset demonstrates that $k$ can adopt negative values for sufficiently large Yukawa interaction strengths.

We point out that the perturbation results in Eqs.~\ref{lp98} and \ref{gt54} (the broken lines in Figs.~\ref{fig1}-\ref{fig3}) provide a good fit of the full result for $F_0/A$, $k c_0$, and $k$ according to Eq.~\ref{lp08} in the region $0 <\bar{a}/\mbox{nm} \lesssim 0.5$, which we expect to be the most relevant range for small ions in aqueous solution \cite{kalcher09,kalcher10}. Note that the broken lines in Figs.~\ref{fig1}-\ref{fig3} are not straight because the abscissa displays $\bar{a}$ (and not the perturbation parameter $a$). The good fit of the perturbation prediction for any variations of $\bar{a}$, $|\bar{b}|$, and $\bar{c}$ (and combinations thereof) in the range from $0$ to about $0.5 \: \mbox{nm}$ is a general observation; see for example the inset of Fig.~\ref{fig1}.

\subsection{Retaining only Yukawa interactions between ions and surface} \label{ii66}
The third specific case assumes we switch off the Yukawa interactions between pairs of mobile ions but retain the Yukawa interactions between the ions and surface. It is interesting to analyze this case because it allows us to assess the relevance of non-electrostatic ion-surface versus ion-ion interactions. The absence of Yukawa pair interactions between mobile ions translates into replacing ${\cal B}$ in Eq.~\ref{zr39} by
\begin{equation}\label{ji52}
{\cal B}=
\begin{pmatrix}
  1  & -1/2 & 1/2 \\
  0  & \tilde{\kappa}^2 & 0 \\
  0  & 0 & \tilde{\kappa}^2
\end{pmatrix}
\end{equation}
without changing ${\cal M}$. That is, ${\cal M}$ remains specified by Eq.~\ref{fy48}, and $F_0/A$, $k c_0$, $k$, and $\bar{k}$ continue to being calculated through Eq.~\ref{lp08}. With this we find the explicit expressions
\begin{eqnarray} \label{ju41}
  \frac{F_0}{A}&=&\frac{2 \pi}{\kappa} \boldsymbol{\sigma}^T
    \left(\begin{array}{ccc}
    \tilde{\kappa} l_B  & \frac{a-b}{2 g_2} & \frac{b-c}{2 g_2} \\
    0 & a & b \\ 
    0 & b & c 
    \end{array} \right) \boldsymbol{\sigma},\nonumber\\
k c_0&=&\frac{\pi}{\kappa^2} \boldsymbol{\sigma}^T
    \left(\begin{array}{ccc}
    \tilde{\kappa}^2 l_B  & \frac{a-b}{2 g_4} & \frac{b-c}{2 g_4} \\
    0 & a & b \\ 
    0 & b & c
    \end{array} \right) \boldsymbol{\sigma},\\    
k&=&\frac{3}{2} \frac{\pi}{\kappa^3} \boldsymbol{\sigma}^T
    \left(\begin{array}{ccc}
    \tilde{\kappa}^3 l_B  & \frac{a-b}{2 g_6} & \frac{b-c}{2 g_6} \\
    0 & a & b \\ 
    0 & b & c 
    \end{array} \right) \boldsymbol{\sigma},\nonumber
\end{eqnarray} 
where we recall $\tilde{\kappa}=\kappa l_D$ and the definitions $g_2=1+\tilde{\kappa}$, $g_4=1$, and $g_6=(1+\tilde{\kappa})/[1+\tilde{\kappa}+\tilde{\kappa}^2]$, as initially introduced following Eq.~\ref{gt54}. 

If in Eq.~\ref{ju41} we set $a=b=c$ (symmetric hydration interactions) we obtain
\begin{eqnarray} \label{hg641}
\frac{F_0}{A}&=&2 \pi l_B l_D {\left(\frac{\sigma_e}{e}\right)}^2+2 \pi a \frac{(\sigma_a+\sigma_c)^2}{\kappa},\nonumber\\
k c_0&=&\pi l_B l_D^2 {\left(\frac{\sigma_e}{e}\right)}^2+\pi a \frac{(\sigma_a+\sigma_c)^2}{\kappa^2},\\
k&=&\frac{3}{2} \pi l_B l_D^3 {\left(\frac{\sigma_e}{e}\right)}^2+\frac{3}{2} \pi a \frac{(\sigma_a+\sigma_c)^2}{\kappa^3}.\nonumber
\end{eqnarray} 
The Yukawa contributions to these results (the second of the two contributions to the right-hand side of Eq.~\ref{hg641}) can be rationalized by the same argument as that leading to the integration in Eq.~\ref{pw91}: particles that are uniformly distributed on a surface with a combined area density $\sigma_a+\sigma_c$ exhibit mutual Yukawa interactions $a e^{-\kappa_{\mathrm{eff}} r}/r$. Yet, in the present case $\kappa_{\mathrm{eff}}=\kappa$ because no Yukawa interactions between the salt ions are present. Because switching off the Yukawa interactions between the mobile salt ions increases the effective characteristic screening length from $1/\sqrt{\kappa^2+2/l_a^2}$ to $1/\kappa$, the free energy (that is, all the quantities $F_0$, $k c_0$, and $k$) increases too. Hence, for any choice $\sigma_a+\sigma_c \neq 0$, the predictions of Eq.~\ref{hg64} for $F_0/A$, $k c_0$, and $k$ are smaller than the corresponding values in Eq.~\ref{hg641}. This explains the somewhat unexpected result that adding Yukawa ion-ion repulsion in addition to Yukawa ion-surface interactions always {\em decreases} $F_0/A$, $k c_0$, and $k$ if $a=b=c$.

\begin{figure}[ht!]
\begin{center}
\includegraphics[width=8.0cm]{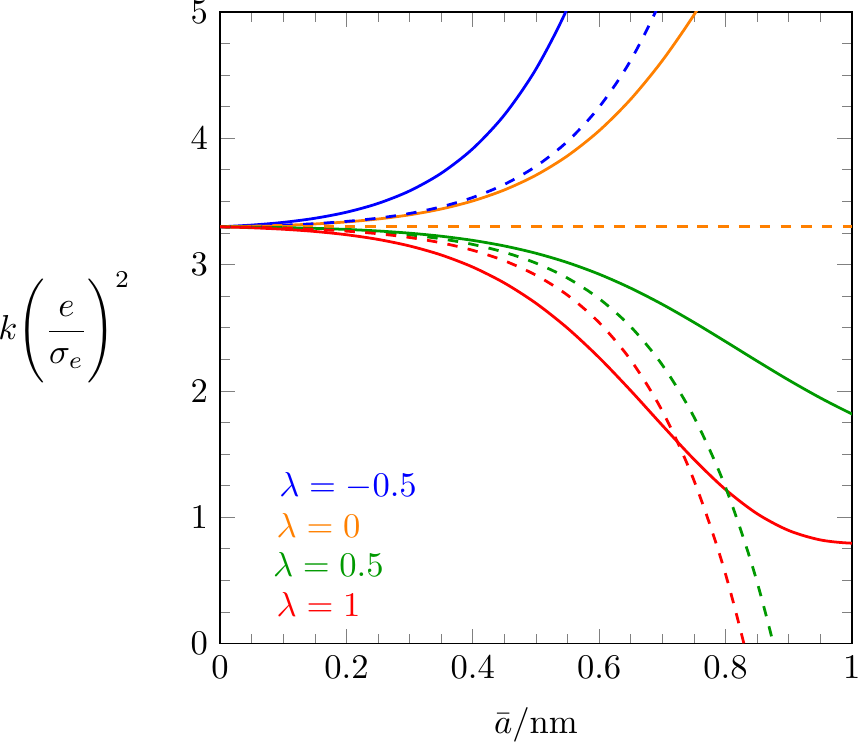}
\end{center}
\caption{\label{fig4} 
Scaled bending stiffness, $k \times (e/\sigma_e)^2$ (in units of $k_BT \mbox{nm}^4$), of a planar surface as function of $\bar{a}$ for $l_B=0.7 \: \mbox{nm}$, $l_D=1 \: \mbox{nm}$, $1/\kappa=0.2 \: \mbox{nm}$, $\bar{b}=\bar{c}=0$. Solid lines refer to the full result in Eq.~\ref{lp08}, broken lines to the result in Eq.~\ref{ju41}. Curves of different color correspond to different couplings, $\chi_a=\lambda \chi_a^{\mathrm{opt}}$ and $\chi_c=\lambda \chi_c^{\mathrm{opt}}$, with $\chi_c^{\mathrm{opt}}=-\chi_a^{\mathrm{opt}}=1/(4 g_6)=1.29$, and $\lambda=-0.5$ (blue), $\lambda=0$ (orange, vanishing coupling), $\lambda=0.5$ (green), $\lambda=1$ (red, optimal coupling).}
\end{figure}

In the case of asymmetric hydration interactions in Eq.~\ref{ju41} (thus allowing for general choices of $a$, $b$, $c$ with $a \ge 0$ and $c \ge 0$) we introduce, as before, coupling parameters $\chi_a$ and $\chi_c$ through $\sigma_a=\chi_a \sigma_e/e$ and $\sigma_c=\chi_c \sigma_e/e$. We obtain optimal coupling parameters $\chi_c^{\mathrm{opt}}=-\chi_a^{\mathrm{opt}}=1/(4 g_2)$ for $F_0/A$, $\chi_c^{\mathrm{opt}}=-\chi_a^{\mathrm{opt}}=1/(4 g_4)$ for $k c_0$, and $\chi_c^{\mathrm{opt}}=-\chi_a^{\mathrm{opt}}=1/(4 g_6)$ for $k$. Inserting these into their corresponding expressions, $F_0/A$, $k c_0$, and $k$, results in 
\begin{eqnarray} \label{fe41}
\frac{F_0}{A}&=&2 \pi l_B l_D {\left(\frac{\sigma_e}{e}\right)}^2 \left[1-\frac{a-2 b+c}{16 l_B g_2^2 \tilde{\kappa}}\right],\nonumber\\
k c_0&=&\pi l_B l_D^2 {\left(\frac{\sigma_e}{e}\right)}^2 \left[1-\frac{a-2 b+c}{16 l_B g_4^2 \tilde{\kappa}^2}\right],\\
k&=&\frac{3}{2} \pi l_B l_D^3 {\left(\frac{\sigma_e}{e}\right)}^2 \left[1-\frac{a-2 b+c}{16 l_B g_6^2 \tilde{\kappa}^3}\right]\nonumber
\end{eqnarray}
at optimal coupling (that is, fixed surface potentials $\Psi_a(x=0)=\Psi_c(x=0)=0$). Hence, in the presence of only Yukawa ion-surface interactions, $F_0/A$ turns negative for $a-2 b+c>16 l_B g_2^2 \tilde{\kappa}$, $k c_0$ turns negative for $a-2 b+c>16 l_B g_4^2 \tilde{\kappa}^2$, and  $k$ turns negative for $a-2 b+c>16 l_B g_6^2 \tilde{\kappa}^3$. Fig.~\ref{fig4} illustrates this for the case of the bending stiffness. The figure shows $k \times (e/\sigma_e)^2$ according to Eq.~\ref{lp08} (solid lines, the full result) and Eq.~\ref{ju41} (broken lines, signifying the absence of Yukawa ion-ion interactions) for the same parameters as in Figs.~\ref{fig1}-\ref{fig3}. 
The result for $k$ in Eq.~\ref{fe41} is displayed by the red broken line. It predicts $k=0$ for $a=16 l_B g_6^2 \tilde{\kappa}^3=54.4 \: \mbox{nm}$ or, equivalently, $\bar{a}=0.83 \: \mbox{nm}$. Hence, for $l_B=0.7 \: \mbox{nm}$, $l_D=1 \: \mbox{nm}$, $1/\kappa=0.2 \: \mbox{nm}$, and $\bar{b}=\bar{c}=0$, the smallest value of $\bar{a}$ for which $k$ may decrease to a vanishing value is $\bar{a}=0.83 \: \mbox{nm}$.

To discuss the physical reason of how the bending stiffness $k$ can adopt negative values, we assume $b=c=0$. The expression for $k$ in Eq.~\ref{ju41} then reads
\begin{equation}
k=\frac{3}{2} \pi l_D^3 \left[l_B \left(\frac{\sigma_e}{e}\right)^2+\frac{a}{\tilde{\kappa}^3} \sigma_a^2+\frac{a}{2 \tilde{\kappa}^3 g_6} \frac{\sigma_e}{e} \sigma_a\right],
\end{equation}
which  immediately reveals the condition $a>16 l_B g_6^2 \tilde{\kappa}^3$ for $k<0$ as stated above. The bending stiffness will be positive when $\sigma_e$ or $\sigma_a$ are increased individually. Negative bending stiffness reflects the coupling between $\sigma_e$ and $\sigma_a$. That is, when $\sigma_e$ is increased from zero to a positive value, anions accumulate in the vicinity of the macroion. These anions experience an additional attraction to the macroion surface due to their Yukawa interaction with the surface. When this additional attraction is strong enough, it renders $k$ negative. We finally note that while subsection \ref{ii66} does not consider Yukawa ion-ion interactions, the bending stiffness can become negative even when Yukawa ion-ion interactions are accounted for; see for example the inset of Fig.~\ref{fig3}. 

\section{Concluding Remarks}
Eq.~\ref{lp08}, together with its derivation and discussion, is the principal outcome of the present work. It specifies the free energy of an isolated, weakly curved macroion in a symmetric 1:1 electrolyte, in the presence of a solvent such as water, on the level of linear Debye-H\"uckel theory, thereby accounting for a composite Coulomb-Yukawa interaction potential among the ions and between the ions and the macroion surface. The curvature-dependent contributions to the free energy are expressed through familiar curvature elastic constants: the bending stiffness, the Gaussian modulus, and the spontaneous curvature. The Yukawa interactions, which embody ion specific effects through a number of independent parameters ($a$, $b$, and $c$, for anion-anion, anion-cation, and cation-cation interactions, respectively, as well as $\sigma_a$ and $\sigma_c$ for surface-mediated interactions with anions and cations), increase the complexity of the Debye-H\"uckel model significantly, despite its linearity. At the same time, the predicted behavior for bending stiffness and spontaneous curvature becomes much richer and can promote, or even induce, curvature instabilities. 
A perturbation approach that yields simple analytic expressions for the curvature-dependent free energy provides excellent agreement with the full model within the experimentally most relevant ranges of the interaction parameters $a$, $b$, and $c$. Simple analytic expressions are also obtained in the limit of switching off the Yukawa interactions among the ions while retaining the Yukawa ion-surface interaction.

Burak and Andelman \cite{burak_2000,burak01} have recently presented a modeling approach that bears some similarity with our present work. They add a short-range, non-electrostatic, hydration-mediated component to the Coulomb pair potential and treat it on the basis of a virial expansion up to lowest order so that their free energy amounts to setting the direct correlation function equal to the pair interaction potential and all higher order direct correlation functions to zero. Hence, while Burak and Andelman \cite{burak01} account for correlations due to a short-range potential on the lowest possible order, our approach completely ignores correlations. The main advantage of our approach, however, is its mathematical simplicity, which originates from the introduction of the auxiliary fields $\Psi_a$ and $\Psi_c$ and which is what allows us to derive simple analytic expressions for the free energy of a weakly curved macroion.

Our target in the present work has been the linear Debye-H\"uckel limit, but that should ultimately be extended to the nonlinear theory based on Eqs.~\ref{er43}. Another future improvement of our model should allow for dielectric inhomogeneities. Our present work assumes a uniform dielectric background, characterized by a constant Bjerrum length of $l_B=0.7$ nm. However, hydration-mediated non-electrostatic ion-ion interactions originate in the ordering of water molecules around each ion, which affects the local dielectric constant. Methods to account for dielectric inhomogeneities \cite{ben11b}, including the Dipolar Poisson-Boltzmann theory that accounts for solvent molecules explicitly as Langevin dipoles \cite{abrashkin07,levy12} are available, but the connection between the explicit account of the solvent and effective hydration-mediated ion-ion interactions is not obvious.

\begin{acknowledgments} 
GVB acknowledges a doctoral scholarship from CAPES Foundation/Brazil Ministry of Education (Grant No.~9466/13-4).
\end{acknowledgments}

\section*{Appendix I: Free Energy Minimization}
We have defined the two potentials $\Psi_a$ and $\Psi_c$ in Eq.~\ref{nn2}. An equivalent definition of these potentials at position ${\bf r}$ is
  \begin{equation}\label{pp2}
   \begin{pmatrix}
   \Psi_a({\bf r})  \\ \Psi_c({\bf r})\end{pmatrix}=\int 
\limits  d^3{\bf r}'
\: \frac{e^{-\kappa |{\bf r}-{\bf r}'|}}{|{\bf r}-{\bf r}'|} \:  
\mathcal{A} \begin{pmatrix}
   n_a({\bf r}')-n_0 \\ n_c({\bf r}')-n_0 \end{pmatrix}.
  \end{equation}
  where we recall $n_a({\bf r})$ and $n_c({\bf r})$ are the local anion and cation concentrations, $n_0$ is their bulk value, and the symmetric square matrix ${\cal A}$ is defined in Eq.~\ref{jo48}. Equivalency between Eqs.~\ref{nn2} and \ref{pp2} is established using the Greens function $G({\bf r})=-e^{-\kappa |{\bf r}|}/(4 \pi |{\bf r}|)$ of the equation $(\nabla^2-\kappa^2) G({\bf r})=\delta({\bf r})$, where $\delta({\bf r})$ is the Dirac delta function. We emphasize again that $\Psi_a$ and $\Psi_c$ reflect concentration changes relative to the bulk. As shown in previous work \cite{caetano16}, the mean-field free energy that includes the solvent-mediated hydration contribution based on our Yukawa potentials reads
\begin{eqnarray}\label{lp21}
F&=&\int d^3{\bf r} \Bigg\{
\frac{(\nabla \Psi_e)^2}{8 \pi l_B}+f_{\mathrm{mix}}(n_a)+f_{\mathrm{mix}}(n_c)
\\
+\frac{1}{8 \pi} &\Bigg[& \begin{pmatrix}  \nabla \Psi_a \\ 
\nabla \Psi_c \end{pmatrix}^T
\!\!\!\! {\cal A}^{-1}
\begin{pmatrix} \nabla \Psi_a \\ \nabla \Psi_c \end{pmatrix}+
\kappa^2 \begin{pmatrix} \Psi_a \\ \Psi_c \end{pmatrix}^T
\!\!\!\! {\cal A}^{-1}
\begin{pmatrix} \Psi_a \\ \Psi_c \end{pmatrix}
\Bigg]\Bigg\},\nonumber 
\end{eqnarray}
where ${\cal A}^{-1}$ is the inverse of ${\cal A}$ and $f_{mix}(n)=n \ln (n/n_0)-n+n_0$ is the mixing free energy (per volume element) of an ideal gas that has a local concentration $n$ and is in equilibrium with a bulk system of fixed concentration $n_0$. Variation of the free energy leads to the expression
\begin{eqnarray}\label{ko51}
\delta F&=&\int do \left[ \Psi_e \frac{\delta \sigma_e}{e}+\Psi_a \delta \sigma_a+\Psi_c \delta \sigma_c \right]+\nonumber\\
&+& \int \limits  d^3{\bf r} \left[\delta n_a \left(-\Psi_e+\Psi_a+\ln \frac{n_a}{n_0}\right)\right]+\\
&+& \int \limits  d^3{\bf r} \left[\delta n_c \left(\Psi_e+\Psi_c+\ln \frac{n_c}{n_0}\right)\right]. \nonumber
\end{eqnarray}
The integration in the first line of Eq.~\ref{ko51} extends over the macroion surface, and the other two integrations run over the volume occupied by the electrolyte. Vanishing of $\delta F$ in thermal equilibrium implies both the Boltzmann distributions in Eq.~\ref{ik32} and the charging free energy 
\begin{equation}\label{hu62}
F=\int do \left[ \int \limits_0^{\sigma_e} \Psi_e \frac{d\bar{\sigma}_e}{e}+\int \limits_0^{\sigma_a}  \Psi_a d\bar{\sigma}_a+\int \limits_0^{\sigma_c} \Psi_c d\bar{\sigma}_c \right],
\end{equation}
where the potentials $\Psi_e$, $\Psi_a$, and $\Psi_c$ are functions of the charging parameters $\bar{\sigma}_e$, $\bar{\sigma}_a$, $\bar{\sigma}_c$ that change from zero to their final values $\sigma_e$, $\sigma_a$, $\sigma_c$, respectively. The order of carrying out these ``charging processes'' is irrelevant. In the linear limit of the Debye-H\"uckel model the potentials $\Psi_e$, $\Psi_a$, and $\Psi_c$ depend linearly on the densities $\bar{\sigma}_e$, $\bar{\sigma}_a$, and $\bar{\sigma}_c$ so that the ``charging process'' can be carried out. The result is Eq.~\ref{hj42}.

\section*{Appendix II: Eigenvalue perturbation theory}
To compute the first-order corrections of the results in Eqs.~\ref{lp98} we denote by ${\cal B}_0$ the matrix ${\cal B}$ with $a=b=c=0$, which possesses eigenvalues
\begin{equation}
    \lambda_1=1,\quad \lambda_2=\widetilde{\kappa}^2,\quad \lambda_3=\widetilde{\kappa}^2
\end{equation}
with corresponding right column eigenvectors
\begin{align}
   & {\bf x}_1=\left(1,0,0\right),\quad {\bf x}_2=\left(\frac12\left(\widetilde{\kappa}^2-1\right)^{-1},0,1\right),\nonumber\\
\quad &{\bf x}_3=\left(-\frac12\left(\widetilde{\kappa}^2-1\right)^{-1},1,0\right), \label{eigenvectors}
\end{align}
satisfying 
\begin{equation} {\cal B}_0{\bf x}_i=\lambda_i {\bf x}_i. \label{eigenvalue_eq}\end{equation}
These eigenvectors are not orthonormal with respect to the standard inner product on $\mathbb{R}^3$. In what follows, it will be convenient to have a quadratic form that renders the eigenvectors orthonormal, i.e.~a symmetric, nondegenerate matrix $Q$ so that ${\bf x}_i^TQ{\bf x}_j=\delta_{ij}.$
If we introduce a matrix $P$ of eigenvectors, this condition is equivalent to $P^TQP=I$, and $Q$ can be computed as $Q=\left(P^T\right)^{-1}P^{-1}=\left(PP^T\right)^{-1}.$

We now consider the perturbation of the defining equation for the eigenvalues and eigenvectors of ${\cal B}_0$. Specifically, we define a matrix $\delta {\cal B}$ by 
the linearization ${\cal B}(\delta a,\delta b,\delta c)\approx{\cal B}_0+\delta {\cal B}$ and consider
\begin{equation} ({\cal B}_0+\delta {\cal B})({\bf x}_i+\delta {\bf x}_i)=(\lambda_i+\delta\lambda_i)({\bf x}_i+\delta {\bf x}_i).   \end{equation}
Making use of the unperturbed equation (Eq.~\ref{eigenvalue_eq}) and keeping terms to first-order, we find
\begin{equation}{\cal B}_0(\delta {\bf x}_i)+(\delta {\cal B}) {\bf x}_i=\lambda_i(\delta {\bf x}_i)+(\delta \lambda_i) {\bf x}_i.       \label{first_order}      \end{equation}
We can expand each perturbation $\delta {\bf x}_i$ in the eigenbasis specified in Eq.~\ref{eigenvectors}, 
\begin{equation}\delta {\bf x_i}=\sum\limits_{j=1}^3 c_{ij}{\bf x}_j.      
\end{equation}
Substituting this into Eq.~\ref{first_order} and using Eq.~\ref{eigenvalue_eq} leads to two cases. When $i=k$, we obtain the eigenvalue perturbations
\begin{equation}  \delta\lambda_i={\bf x}_i^TQ(\delta {\cal B}) {\bf x}_i.            \end{equation}
Otherwise, $i\neq k$ and we make the replacement $k\rightarrow j$ and calculate the expansion coefficients
\begin{equation} c_{ij}=\frac{{\bf x}_j^TQ(\delta {\cal B}) {\bf x}_i}{\lambda_i-\lambda_j},\quad i\neq j.            \end{equation}
The coefficients $c_{ii}$, $c_{23}$, and $c_{32}$ cannot be calculated unless an additional constraint is imposed on the eigenvectors, but we will see that our results are independent of these quantities.

Calculation of all physical quantities of interest consists essentially in computing $({\cal B}_0+\delta {\cal B})^r$. Non-integer powers of matrices are defined through diagonalization:
$ {\cal B}_0^r=P\Lambda^r P^{-1}, $
where $\Lambda=\mathrm{diag}(\lambda_1,\lambda_2,\lambda_3)$ and $\Lambda^r=\mathrm{diag}(\lambda_1^r,\lambda_2^r,\lambda_3^r)$. There is an issue of uniqueness when $r$ is non-integral; as our eigenvalues are real and positive, this is resolved by taking the positive branch of each quantity $\lambda_i^r$. We consider the perturbation
\begin{equation} ({\cal B}_0+\delta {\cal B})^r=(P+\delta P)(\Lambda+\delta\Lambda)^r( P+\delta P)^{-1}   \label{matrix_power_perturbation}  \end{equation}
where $\delta P$ consists of the eigenvector perturbations $\delta {\bf x}_i$. When the perturbations are sufficiently small, the second and third factors on the right-hand side of Eq.~\ref{matrix_power_perturbation} can be calculated as binomial expansions, leading to
\begin{equation}
    ({\cal B}_0+\delta {\cal B})^r={\cal B}_0^r+rP(\delta\Lambda) P^{-1}+\left[(\delta P)P^{-1},{\cal B}_0^r \right],
\end{equation}
to first order and where $[\cdot,\cdot]$ is the matrix commutator. It is easily verified (using a simultaneous diagonalization argument) that this result is independent of the quantities $c_{ii}$, $c_{23}$, and $c_{32}$. 

%

\end{document}